\documentclass[prx,twocolumn,superscriptaddress]{revtex4-1}

\usepackage{amsmath,amssymb,amsthm}
\usepackage{bm}
\usepackage{bbm}
\usepackage{array}
\usepackage{enumerate}
\usepackage{amsfonts}
\usepackage{latexsym,epsfig,epstopdf}
\usepackage{graphicx}

\usepackage[UKenglish]{babel}
\usepackage{appendix}
\usepackage{hyperref}
\usepackage{cleveref}

\makeatletter
\DeclareRobustCommand{\cev}[1]{%
  \mathpalette\do@cev{#1}%
}
\newcommand{\do@cev}[2]{%
  \fix@cev{#1}{+}%
  \reflectbox{$\m@th#1\vec{\reflectbox{$\fix@cev{#1}{-}\m@th#1#2\fix@cev{#1}{+}$}}$}%
  \fix@cev{#1}{-}%
}
\newcommand{\fix@cev}[2]{%
  \ifx#1\displaystyle
    \mkern#23mu
  \else
    \ifx#1\textstyle
      \mkern#23mu
    \else
      \ifx#1\scriptstyle
        \mkern#22mu
      \else
        \mkern#22mu
      \fi
    \fi
  \fi
}

\newcommand{\past}[1]{\cev{#1}}
\newcommand{\future}[1]{\vec{#1}}

\makeatletter
\newcommand*{\balancecolsandclearpage}{%
  \close@column@grid
  \clearpage
  \twocolumngrid
}
\makeatother

\newcommand{\ket}[1]{\left | #1 \right\rangle}
\newcommand{\bra}[1]{\left \langle #1 \right |}
\newcommand{\braket}[2]{\left\langle #1|#2\right\rangle}

\newcommand{\cphase}[0]{C^{\varphi}_q}
\newcommand{\dphase}[0]{D^{\varphi}_q}
\newcommand{\sigmaphase}[1]{\left | \sigma ^{\varphi}_{#1} \right\rangle}
\newcommand{\uphase}[0]{U^{\varphi}}
\newcommand{\cphasemin}[0]{{C^{\varphi}_q}_{\min}}
\newcommand{\dphasemin}[0]{{D^{\varphi}_q}_{\min}}
\newcommand{\brasigmaphase}[1]{\left \langle \sigma ^{\varphi}_{#1} \right |}
\newcommand{\braketsigmaphase}[2]{\left\langle \sigma ^{\varphi}_{#1} | \sigma ^{\varphi}_{#2} \right\rangle }

\newcommand{\onephase}[1]{\left | 1 ^{\varphi}_{#1} \right\rangle}

\newcommand{\braketonephase}[2]{\left\langle 1 ^{\varphi}_{#1} | 1 ^{\varphi}_{#2} \right\rangle }
\newcommand{\cijphase}[2]{{c^{\varphi}_{#1#2}}}
\newcommand{\rhophase}{{\rho^{\varphi}}}

\newcommand{\inlineheading}[1]{\textbf{{#1}}}

\usepackage{color}
\usepackage[dvipsnames]{xcolor}
\usepackage[normalem]{ulem}
\usepackage{xspace}

\newcommand{\figref}[1]{Fig.~\ref{#1}}

\definecolor{blue}{rgb}{0,0.2,1}

\definecolor{red}{rgb}{0.9,0,0}

\newcommand{\Tr}{\mathrm{Tr}}

\newcommand{\CQT}{Centre for Quantum~Technologies, National~University~of~Singapore, 3 Science Drive 2, 117543, Singapore}
\newcommand{\NTUPhys}{School of Physical and Mathematical Sciences, Nanyang Technological University, 637371, Singapore}
\newcommand{\ComI}{Complexity Institute, Nanyang Technological University, 637335, Singapore}

\begin{document}

\title{Optimal stochastic modelling with unitary quantum dynamics}

\author{Qing Liu}
\email{liuqingppk@gmail.com}
\affiliation{\NTUPhys}
\affiliation{\ComI}

\author{Thomas J. Elliott}
\email{physics@tjelliott.net}
\affiliation{\ComI}
\affiliation{\NTUPhys}

\author{Felix C. Binder}
\email{quantum@felix-binder.net}
\affiliation{\NTUPhys}
\affiliation{\ComI}

\author{Carlo Di Franco}
\affiliation{\NTUPhys}
\affiliation{\ComI}

\author{Mile Gu}
\email{gumile@ntu.edu.sg}
\affiliation{\NTUPhys}
\affiliation{\CQT}
\affiliation{\ComI}

\date{\today}

\begin{abstract}
Identifying and extracting the past information relevant to the future behaviour of stochastic processes is a central task in the quantitative sciences. Quantum models offer a promising approach to this, allowing for accurate simulation of future trajectories whilst using less past information than any classical counterpart. Here we introduce a class of phase-enhanced quantum models, representing the most general means of causal simulation with a unitary quantum circuit. We show that the resulting constructions can display advantages over previous state-of-art methods -- both in the amount of information they need to store about the past, and in the minimal memory dimension they require to store this information.  Moreover, we find that these two features are generally competing factors in optimisation -- leading to an ambiguity in what constitutes the optimal model -- a phenomenon that does not manifest classically. Our results thus simultaneously offer new quantum advantages for stochastic simulation, and illustrate further qualitative differences in behaviour between classical and quantum notions of complexity.
\end{abstract}

\maketitle
\enlargethispage{\baselineskip}

Models of stochastic processes are essential to quantitative science, providing a systematic means for simulating future behaviour based on past observations. Given different models exhibiting statistically identical behaviour, there is a general preference for the simplest models -- those which require minimal information about the past. The motivation is two-fold: foundationally, they represent a way of identifying potential causes of future events; and operationally, simulating a process using such models requires less memory -- as they need to track less information about the past -- leading to a reduction in resource costs.

The field of \emph{computational mechanics}~\cite{crutchfield1989inferring, shalizi2001computational, crutchfield2012between} provides a systematic approach to constructing the provably simplest classical causal model for any given stochastic process. These models, called $\varepsilon$-machines, can produce statistically correct predictions using less memory than any classical alternative. The amount of past information they store has been employed as a measure of structure in diverse contexts~\cite{crutchfield1997statistical, palmer2000complexity, varn2002discovering, clarke2003application, park2007complexity, li2008multiscale, haslinger2010computational, kelly2012new}, motivated by its interpretation as a fundamental limit on how much information from the past must be tracked in order to predict the future.

Quantum mechanics, however, enables even simpler models that bear statistically identical predictions~\cite{gu2012quantum, mahoney2016occam, thompson2017using, aghamohammadi2018extreme, binder2018practical, elliott2018superior, elliott2018quantum, riechers2016minimized,yang2018matrix}. This advantage, which has been observed experimentally~\cite{palsson2017experimentally, ghafari2017observing}, can scale without bound~\cite{garner2017provably, aghamohammadi2017extreme, elliott2018superior, thompson2018causal} and induces significant qualitative classical-quantum divergences in quantifiers of structure~\cite{suen2017classical,aghamohammadi2016ambiguity}. However, while presently-known quantum constructions are provably optimal for some specific cases~\cite{suen2017classical, thompson2018causal}, they are known not to be so in general. This motivates the search for even simpler quantum models that obtain further memory advantages in stochastic simulation, and better characterise quantum notions of structure and complexity.

In this paper, we introduce \emph{phase-enhanced} quantum models -- a sophistication of previous quantum models -- that capture all possible methods of causal simulation using unitary quantum circuits. We show that the resulting models can improve upon current state-of-the-art constructions in further reducing the amount of memory they require, according to both entropic and dimensional measures~\cite{thompson2018causal}. Moreover, our new models reveal the origin and highlight the widespread nature of a recently discovered phenomenon~\cite{loomis2018strong} -- which we term the \emph{ambiguity of optimality} -- wherein optimising for quantum models that track minimal information about the past may sacrifice achieving minimal dimensionality of their memory (and vice versa).

\section*{Framework}
\inlineheading{Classical models.}
A bi-infinite discrete-time, discrete-event stochastic process~\cite{khintchine1934korrelationstheorie} is characterised by a sequence of random variables $X_t$ that take values $x_t$ drawn from a finite alphabet $\mathcal{A}$ at each time step $t\in \mathbb{Z}$. The process is defined by a joint probability distribution $P(\past{X},\future{X})$, where $\past{X}=\cdots X_{-2} X_{-1}$ and $\future{X}=X_0 X_1 \cdots$ represent the past and future sequences of the process respectively (we use upper case to denote random variables, and lower case for their variates). A consecutive sequence of length $L$ is denoted by $X_{0:L}=X_0\cdots X_{L-1}$. Here, we consider stationary stochastic processes, such that $P(X_{0:L})=P(X_{m:m+L})\forall L,m\in\mathbb{Z}$.

An instance of a given stochastic process has a specific past $\past{x}$, and possesses a corresponding conditional future $P(\future{X}|\past{x})$. A causal model of a stochastic process defines an encoding function that maps each possible $\past{x}$ to some suitable memory state such that the same systematic action on the memory at each timestep gives rise to future sequences according to this conditional future distribution. Notably, all information about the future that is stored in the memory states may be obtained from observations of the past~\cite{crutchfield1989inferring,shalizi2001computational,thompson2018causal}.

The field of computational mechanics~\cite{crutchfield1989inferring, shalizi2001computational} offers a systematic means to construct the simplest classical causal models -- $\varepsilon$-\emph{machines}. These models are defined by encoding past information into \emph{causal states} $s\in\mathcal{S}$, defined by an equivalence relation on the past-future conditional distribution:
\begin{equation}
\past{x},\past{x}'\in s \Leftrightarrow P(\future{X}|\past{x})=P(\future{X}|\past{x}').
\end{equation}
A key property of $\varepsilon$-machines is that they are unifilar~\cite{shalizi2001computational} -- given an initial causal state $s$ and output symbol $x$, the memory transitions into a unique subsequent causal state. We may thus define an update rule $\lambda(s,x)$ to describe the new state~\cite{binder2018practical}.

The memory of an $\varepsilon$-machine is often parameterised according to two metrics~\cite{crutchfield1989inferring}: the \emph{statistical complexity}
\begin{equation}\label{eq:cmu}
C_\mu  := H(S)=-\sum_{s\in S}\pi_s\log_2 \left({\pi_s}\right)
\end{equation}
which measures the amount of information stored in the memory, and the \emph{topological complexity}
\begin{equation}\label{eq:dmu}
D_\mu := \log_2 \left(\mathrm{dim}\left(\mathcal{S}\right)\right)
\end{equation}
which measures the dimension of the memory. Here, $\pi_s = \sum_{\past{x}\in s}P(\past{x})$ denotes the steady-state distribution of the causal states. The $\varepsilon$-machine minimises both these metrics over analogous measures for the memory of all other classical causal models. Nevertheless, it still stores information that is not directly relevant for simulating future statistics; $C_\mu$ can be strictly greater than the mutual information between past and future~\cite{shalizi2001computational}. Operationally, $C_\mu$ and $D_\mu$ correspond to the size of the simulator memory (per simulator), when run in an ensemble or single-shot setting respectively.

\begin{figure}
\begin{center}
\includegraphics[width=0.98\linewidth]{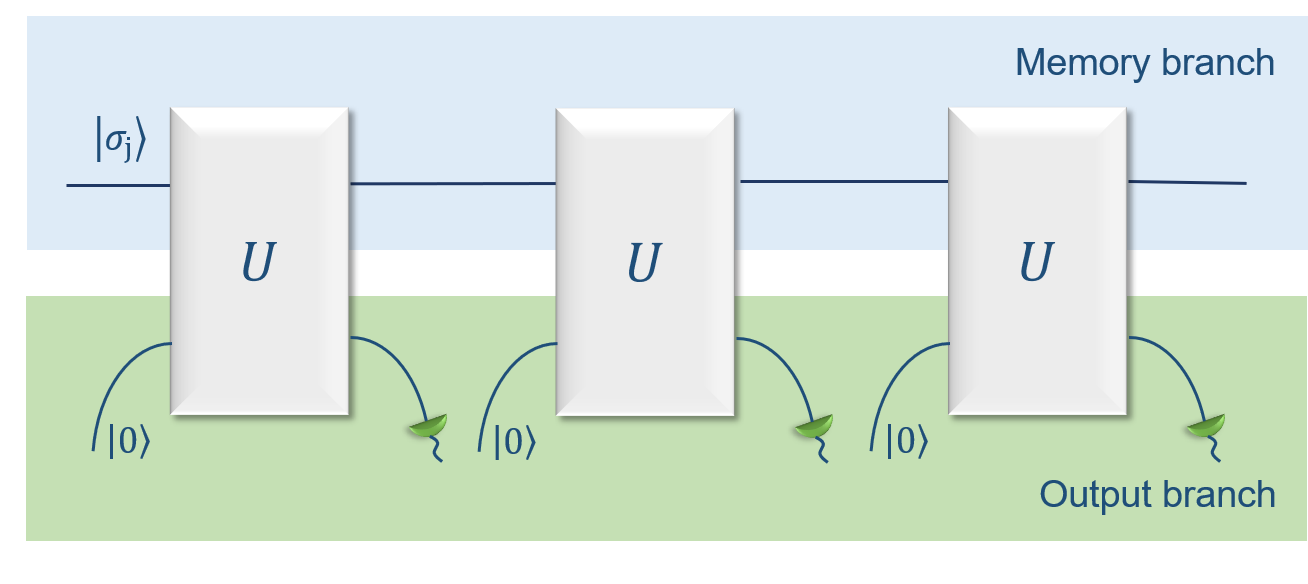}
\end{center}
\caption{A unitary quantum model produces a statistical sequence of outputs $\future{x}=x_0x_1x_2\cdots$ by interacting a blank ancilla with the memory through $U$ at each time step, and then measuring the state of the output branch in the computational basis. }
\label{fig:unitarycircuit}
\end{figure}

\inlineheading{Quantum models.}
Quantum effects present an opportunity to bypass classical limits, enabling models that require less past information than $\varepsilon$-machines~\cite{gu2012quantum, mahoney2016occam, thompson2017using, aghamohammadi2018extreme, binder2018practical, elliott2018superior, elliott2018quantum}. The present state-of-the-art systematic constructions for quantum models can be expressed as a step-wise unitary circuit~\cite{aghamohammadi2018extreme, binder2018practical}, where each causal state $j\in S$ is assigned to a corresponding quantum memory state $\ket{\sigma_j}$.  Future sequences are manifest via the use of a unitary operator $U$ that satisfies
\begin{equation}\label{eq:qsimulator}
U\ket{\sigma_j}\ket{0}=\sum_{x}\sqrt{P(x|j)}\ket{\sigma_{\lambda(j,x)}}\ket{x},
\end{equation}
where we have introduced the shorthand notation $P(x|j)\equiv P(X_{t+1}=x|S_t=j)$. At each time step $t$, the memory state (first subspace) is interacted with a fresh ancilla (second subspace) initialised in $\ket{0}$ [\figref{fig:unitarycircuit}]. Subsequent measurement of the resulting ancilla then yields the correct conditional future statistics at each time step. Such a unitary operation has been proven to exist for any stationary stochastic process~\cite{binder2018practical}.

We can extend the definitions of Eqs.~\eqref{eq:cmu} and~\eqref{eq:dmu} to the quantum domain:
\begin{align}
\label{eq:cdq}
C_q & := -\Tr\left(\rho\log_2\left( \rho\right)\right)\nonumber \\
D_q & := \log_2\left(\mathrm{rank}\left(\rho\right)\right),
\end{align}
where $\rho=\sum_j\pi_j \ket{\sigma_j}\bra{\sigma_j}$. These quantities inherit the same operational significance as their corresponding classical counterparts. We refer to them as the \emph{quantum statistical memory} and \emph{quantum topological memory}, respectively. These quantities are model-dependent~\footnote{Note that $C_q$ has sometimes been alternatively referred to as the quantum statistical complexity or quantum machine complexity; we avoid such nomenclature here as the former is incorrect if the model is not minimal, while the latter invites potential confusion with $D_q$.}.

As the memory states are generally not mutually orthogonal they enable memory savings in terms of both metrics~\cite{nielsen2000quantum}. In fact, the above constructions saturate bounds on pairwise memory state overlap~\cite{suen2017classical, binder2018practical}. That is, for any quantum model the overlap between quantum memory states $c_{jk}:=\braket{\sigma_j}{\sigma_k}$ cannot exceed the fidelities of their respective conditional future distributions $f_{jk}=\sum_{\future{x}}\sqrt{P(\future{x}|j)P(\future{x}|k)}$ due to information processing inequalities. For the above construction, $c_{jk} = f_{jk}$~\cite{binder2018practical,mahoney2016occam}.

Despite this, the optimality of these models is only proven for specific processes~\cite{suen2017classical,thompson2018causal}, and known not be true in general. $C_q$ and $D_q$ are thus not the true quantum analogues of statistical and topological complexity, but rather bound them from above. There is hence a strong motivation to find quantum models whose memories further reduce these measures, in order to both provide a more efficient means of stochastic modelling, and to capture the ultimate limits of quantum models.

\section*{Results}
\inlineheading{Phase-enhanced quantum models.}
We construct our phase-enhanced unitary models by postulating a new set of quantum memory states $\{\sigmaphase{j}\}$ with a corresponding unitary interaction $\uphase$ satisfying a generalisation of Eq.~\eqref{eq:qsimulator}:
\begin{equation}\label{eq:phaseqsimulator}
\uphase \sigmaphase{j} \ket{0} := \sum_{x} \sqrt{P(x|j)} e^{i\varphi_{xj}}\sigmaphase{\lambda(j,x)} \ket{x},
\end{equation}
where $\{e^{i\varphi_{xj}}\}$ are the additional phase factors that depend both on the initial causal state $j$ and the output symbol $x$. Given a set of memory states and unitary operator satisfying this relation, measurements of the second subspace in the computational basis $\{\ket{x}\}$ are guaranteed to produce sequences that obey the same statistics as the corresponding non-phase-enhanced model.

\noindent \textbf{Theorem 1}: \emph{All phase-enhanced models are valid; a corresponding unitary $\uphase$ satisfying Eq. (\ref{eq:phaseqsimulator}) exists for any choice of phase factors $\{e^{i\varphi_{xj}}\}$.}

The proof is given in the Supplementary Material.

\noindent \textbf{Theorem 2}: \emph{The set of phase-enhanced models of a given stochastic process as described above contains the unitary quantum models of the process that minimise each of the quantum statistical and topological memories.}

The only possible valid modifications that can be made to Eq.~\eqref{eq:phaseqsimulator} are refinements~\cite{shalizi2001computational} of the memory states beyond the causal states. Modifying the transition structure between the memory states in any other manner, or modifying the magnitude of the terms in the action of the unitary will change the output statistics, and hence change the process being modelled, ruling out such modifications. It has previously been shown that such refinements can only increase the statistical memory~\cite{suen2017classical}; thus, the minimal unitary quantum models must be described by Eq.~\eqref{eq:phaseqsimulator}

As the quantum memory state overlaps $\cijphase{j}{k}:=\braketsigmaphase{j}{k}$ generally differ between different phase choices, the corresponding memory measures will also differ. For any phase-enhanced model we can compute the corresponding quantum statistical and topological memories:
\begin{align}
\label{eq:cdq}
\cphase & := -\Tr\left(\rhophase\log_2\left(\rhophase\right)\right)\nonumber \\
\dphase & := \log_2\left(\mathrm{rank}\left(\rhophase\right)\right),
\end{align}
where $\rhophase=\sum_j\pi_j \sigmaphase{j}\brasigmaphase{j}$. Since these quantities depend on the choice of $\{\varphi_{xj}\}$, we define
\begin{equation}
\cphasemin:=\mathrm{min}_{\{\varphi_{xj}\}}\cphase,
\end{equation}
(and similarly $\dphasemin$) as the minimal quantum statistical (topological) memory over all possible phase-enhancements. Should these quantities be smaller than those without phase-enhancement, i.e.,
\begin{equation}
\cphasemin < C_q,\quad \dphasemin < D_q,
\end{equation}
the resulting phase-enhanced models would be more memory efficient.

The potential for such a memory reduction might at first blush appear counterintuitive. In the Supplementary Material, we show that the overlaps of the quantum memory states are given by
\begin{equation}
\label{eq:overlapphase}
\cijphase{j}{k} =\sum_{\future{x}}  \sqrt{P(\future{x}|j)P(\future{x}|k)}e^{i(\varphi_{\future{x}k}-\varphi_{\future{x}j})},
\end{equation}
where $\varphi_{x_{m:n}j}:=\sum_{t=m}^n\varphi_{x_t\lambda(x_{m:t},j)}$ is shorthand for the multi-step combination of phases. Therefore, $|\cijphase{j}{k}|$ is always maximised when all phase factors are zero. Moreover, for most other choices, $|\cijphase{j}{k}|$ is strictly less than $c_{jk}$; phase factors cannot increase pairwise overlaps between memory states. Nevertheless, as we illustrate in the next section, phase-enhancement can indeed lead to simpler quantum models according to both memory metrics. The possibility to reduce topological memory can be understood as the phase factors creating linear dependencies between the memory states. Meanwhile, its potential to reduce statistical memory nicely illustrates that increasing pair-wise distinguishability between an ensemble of quantum states can sometimes still  reduce higher-order distances between the ensemble that are captured by the von Neumann entropy~\cite{jozsa2000distinguishability}.

\begin{figure}
\begin{center}
\includegraphics[width=0.9\linewidth]{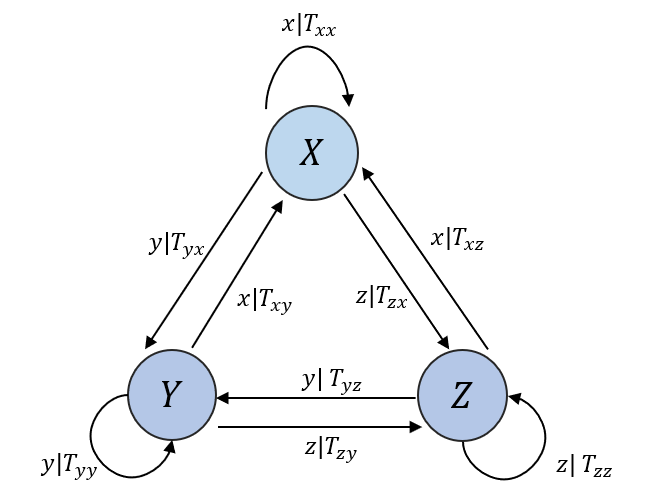}
\end{center}
\caption{General three-state Markov model: The notation $y|T_{yx}$ indicates that the transition from state $x$ to $y$ occurs with probability $T_{yx}$, and the output symbol is $y$.}
\label{fig:general_process}
\end{figure}

\inlineheading{Three-state Markov processes.}
We illustrate the power of phase-enhancements by systematic study of three-state Markovian processes. The Markov chain for  such processes is given in \figref{fig:general_process}, where $T_{yx}$ is used to denote the transition probability of going from state $x$ to state $y$ (while emitting $y$). The Markov property allows us to simplify Eq.~\eqref{eq:phaseqsimulator} to
\begin{equation}
\uphase \sigmaphase{x} \ket{0} = \sum_{y} \sqrt{T_{yx}} e^{i\varphi_{yx}} \sigmaphase{y} \ket{y}.
\end{equation}

\noindent \textbf{Theorem 3}: \emph{Phase-enhancements can reduce the dimension of the memory (i.e., quantum topological memory), providing advantages for single-shot stochastic modelling.}

The condition for dimensional reduction is that there exists a linear dependence between the quantum memory states:
\begin{equation}
\label{eq:lineardep}
\alpha \sigmaphase{x} + \beta \sigmaphase{y} = \sigmaphase{z}
\end{equation}
 for some $\alpha,\beta\in \mathbb{R}^+$. We can restrict $\alpha$ and $\beta$ to be positive reals through freedom to add phase to the memory states $\{\ket{\sigma^\varphi_j}\}$. Moreover, due to global phase invariance, we can set $\varphi_{wx}=0$ for all $w\in\{x,y,z\}$ without loss of generality. Eq.~\eqref{eq:lineardep} can be expressed in terms of the transition probabilities:
\begin{equation}
\alpha \sqrt{T_{wx}}+ \beta \sqrt{T_{wy}} e^{i\varphi_{wy}}=\sqrt{T_{wz}} e^{i\varphi_{wz}}\ \forall\ w.
\end{equation}
From this we obtain the following set of inequalities
\begin{equation}\label{eq:inequality}
\left|\alpha \sqrt{T_{wx}}-\beta \sqrt{T_{wy}}\right|\leq \sqrt{T_{wz}} \leq  \alpha \sqrt{T_{wx}}+\beta \sqrt{T_{wy}}
\end{equation}
that must be satisfied for all $w$. The existence of real and positive $(\alpha,\beta)$ satisfying these inequalities is a necessary and sufficient condition for a dimensional advantage.

Furthermore, given $(\alpha,\beta)$ that satisfy these conditions for a set of transition probabilities, we can determine the phases that collapse the memory to two dimensions:
\begin{align}
\cos(\varphi_{wy}) &= \frac{T_{wz}-\alpha^2T_{wx}-\beta^2T_{wy}}{2\alpha\beta \sqrt{T_{wx}T_{wy}}}\nonumber\\
\cos(\varphi_{wz}) &= \frac{\alpha \sqrt{T_{wx}}+\beta \sqrt{T_{wy}}\cos(\varphi_{wy})}{\sqrt{T_{wz}}}.
\end{align}
Thus, for processes satisfying these inequalities the phase-enhanced quantum model has $\dphasemin=1$, in contrast to the non-phase-enhanced model with $D_q=\log_{2}(3)$.

We performed a numerical sweep over the space of three-state Markov processes (see Supplementary Material), and found that the inequalities Eqs.~\eqref{eq:inequality} are satisfied for approximately $9\%$ of such processes when $\alpha=\beta=1$. Expanding the range of $\alpha$ and $\beta$ values to $\{1,2,3,1/2,1/4\}$ we find that the inequalities can be satisfied by at least $17\%$ of cases. Accounting for additional values for the parameters can only increase this number. However, our lower bound already indicates that dimensional advantages, wherein $\dphasemin<D_q\leq D_\mu$, are relatively commonplace.


\noindent \textbf{Theorem 4}: \emph{Phase-enhancements can reduce the quantum statistical memory.}

\begin{figure}
\begin{center}
\includegraphics[width=\linewidth]{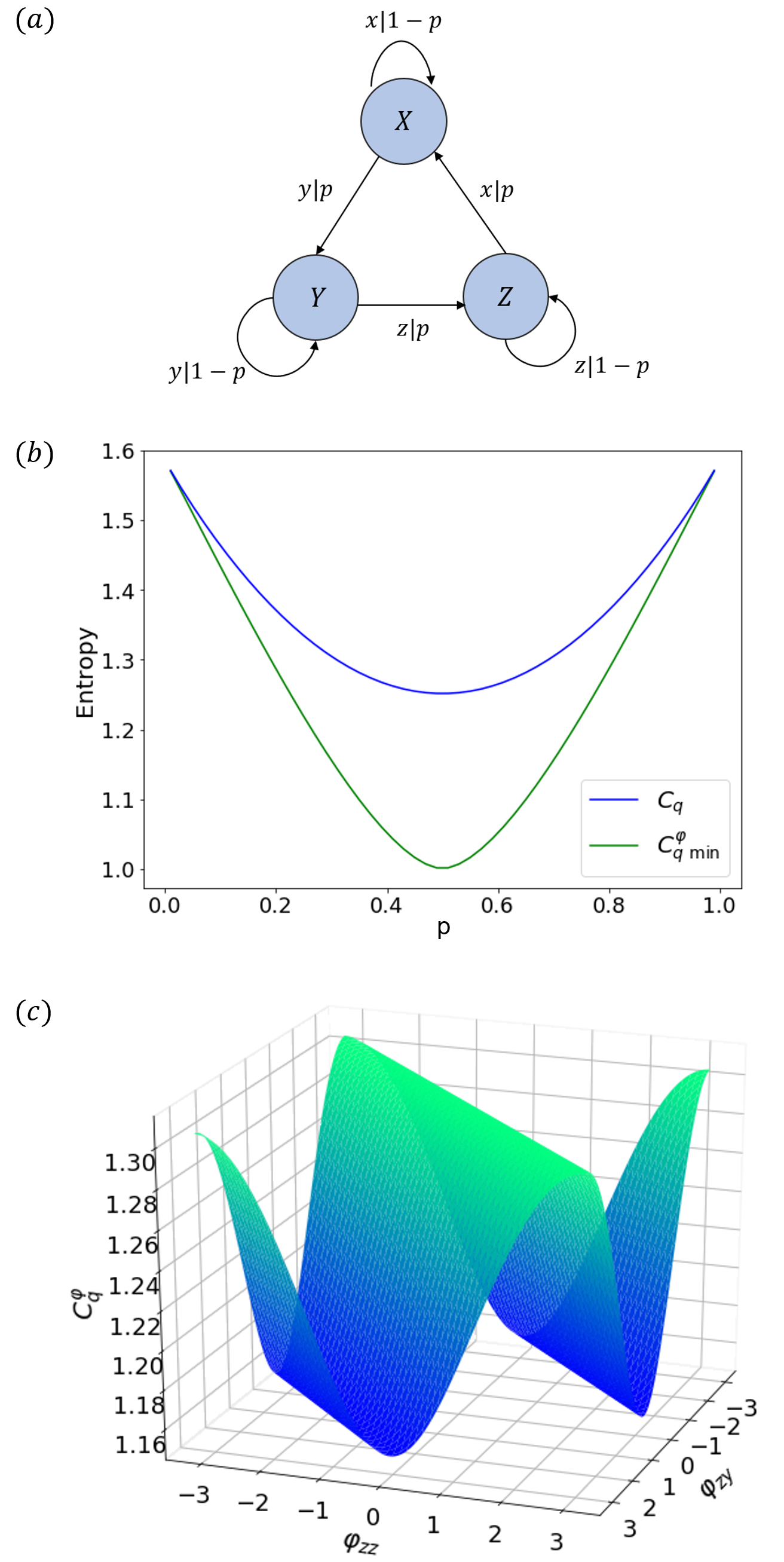}
\end{center}
\caption{(a) Symmetric three-state quasi-cycle. (b) $C_q$ and $\cphasemin$ as a function of $p$. (c) The dependence of $\cphase$ on the phase factors when $p=0.3$.}
\label{fig:ensembleadv}
\end{figure}

Consider a symmetric three-state quasi-cycle \cite{horodecki2013fundamental} as illustrated in \figref{fig:ensembleadv}(a). The transition matrix $T$ for the process is given by
\begin{equation}
T=\begin{pmatrix}
    1-p & 0 & p \\
    p & 1-p & 0 \\
    0 & p & 1-p
\end{pmatrix}.
\end{equation}

Due to certain phase symmetries such as global phase, the quantum memory states $\sigmaphase{j}$ and corresponding unitary $\uphase$ can be given in their most general form as:
\begin{align}
\uphase \sigmaphase{x} \ket{0} & =\sqrt{1-p}\sigmaphase{x}\ket{x} + \sqrt{p}\sigmaphase{y}\ket{y}\nonumber\\
\uphase \sigmaphase{y} \ket{0} & =\sqrt{1-p} \sigmaphase{y}\ket{y}+\sqrt{p}e^{i\varphi_{zy}}\sigmaphase{z}\ket{z}\nonumber\\
\uphase \sigmaphase{z} \ket{0} & =\sqrt{p}\sigmaphase{x}\ket{x}+\sqrt{1-p} e^{i\varphi_{zz}}\sigmaphase{z}\ket{z}.
\end{align}

We calculate the statistical memory $\cphase$ for this model across the full range of possible phase factors. In \figref{fig:ensembleadv}(b) we compare $\cphasemin$ with $C_q$, observing a clear advantage with our phase-enhanced models. We also show the full dependence of $\cphase$ on the two phase parameters in \figref{fig:ensembleadv}(c), where it can be seen that $\cphasemin$ is found when $|\varphi_{zy}-\varphi_{zz}| = \pi$.

Performing a numerical sweep across the space of general three-state Markov processes however, we find that entropic advantages appear to be quite rare, occuring in less than $0.5\%$ of cases (see Supplementary Material).


Our numerical results thus indicate that for three-state Markov processes, models that admit $\dphasemin<D_q$ are much more common than those with $\cphasemin<C_q$. This begets the question, what happens to $\cphase$ for models with dimensional advantages? We find that in many cases for which $\dphasemin<D_q$, the corresponding $\cphase$ is strictly greater than $C_q$. However, since multiple choices of phases can provide a dimensional advantage, one may be tempted to think that another set of phases will show advantages in both metrics. We now study a family of processes that conclusively show that the dichotomy cannot always be resolved in this manner: unlike classical causal models, the optimal quantum model can depend on the choice of memory metric.

\noindent \textbf{Theorem 5}: \emph{The model that minimises quantum topological memory is not in general that which minimises quantum statistical memory. That is, there is no unique optimal quantum model, leading to an \emph{ambiguity of optimality}.}

A process displaying this phenomenon for models with real phases was recently highlighted~\cite{loomis2018strong}. Our results here illustrate that this phenomenon is in fact widespread when general complex phase-enhancements are introduced.

\begin{figure}
\begin{center}
\includegraphics[width=\linewidth]{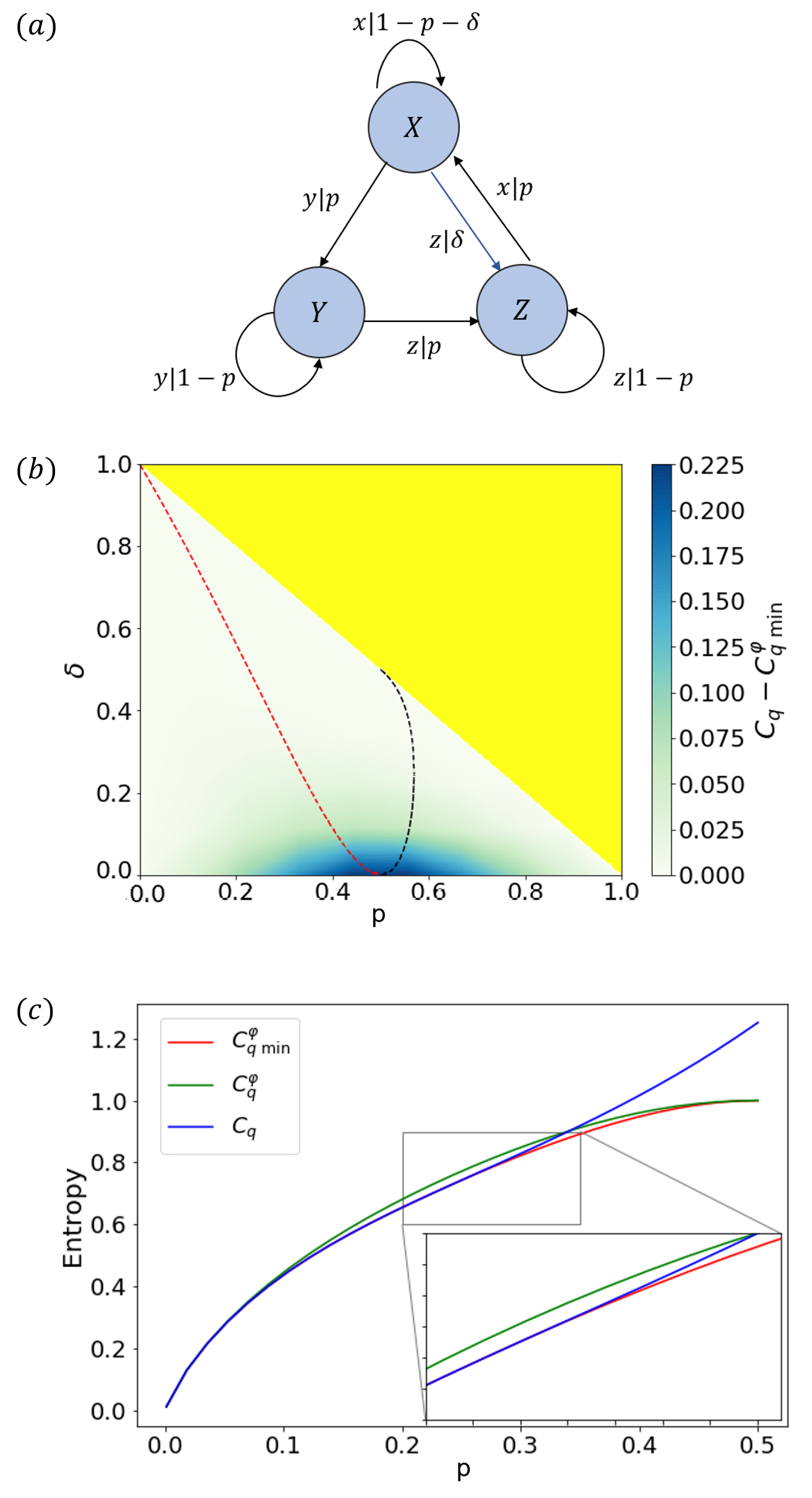}
\end{center}
\caption{(a) Modified three-state quasi-cycle with slippage. In (b) we show the regions with entropic (colour plot) and dimensional (dashed line) advantages. The yellow region delineates the non-physical parameter regime. In (c) we show that the choice of phases that lead to $\cphasemin$ does not always correspond to $\dphasemin$, giving rise to an ambiguity of optimality.}
\label{fig:ambiguity}
\end{figure}

Consider a modified three-state quasi-cycle with slippage, as illustrated in \figref{fig:ambiguity}(a). Our phase-enhanced models offer dimensional advantages along one line of the parameter space, while there is a large area of the space that permits models that exhibit an entropic advantage [\figref{fig:ambiguity}(b)]. Specifically, a dimensional advantage exists iff $p$ and $\delta$ satisfy $\sqrt{p\delta(1-p)}-p\sqrt{p}=(1-p)\sqrt{1-p-\delta}$, in which case the inequalities Eqs.~\eqref{eq:inequality} are satisfied only for a single pair of values of $\alpha$ and $\beta$ given by
\begin{align}
\alpha&=\frac{\sqrt{p}}{\sqrt{1-p-\delta}}\nonumber\\
\beta&=\frac{-p}{\sqrt{(1-p)(1-p-\delta)}}
\end{align}
Since there is only a single set of values for $(\alpha,\beta)$ that offer a linear dependence between the memory states at each point along the aforementioned line, we can be satisfied that this gives the unique optimal model in terms of topological memory. In \figref{fig:ambiguity}(c) we plot $\cphase$ for this model in the parameter region denoted by the red dashed line, and compare it to $C_q$ and $\cphasemin$. We see that for certain parameter values $\cphase>C_q>\cphasemin$, confirming the ambiguity of optimality.

Geometrically, we can understand how such an ambiguity can manifest; reductions in topological memory require linear dependence between the memory states, irrespective of the distance between them, while reductions in statistical memory arise from reductions in the distance between the states. When these two factors are in competition, the ambiguity occurs.

\section*{Discussion}

We have shown that complex phase-based encodings can provide further memory advantages for quantum models of stochastic processes, beyond the previous state-of-the-art constructions. We have provided examples of such enhancements, and through these, demonstrated an ambiguity in which model should be considered optimal based on the measure of memory, a phenomenon not present for classical models. Nevertheless, our resulting models are proven to contain the optimal models among all unitary constructions for each measure of memory.

A natural next step is to explore the prevalence of phase-enhanced models and associated ambiguities in higher dimensions. We expect that such enhancements will become more typical in stochastic processes with larger numbers of causal states -- the rationale being that the number of phase parameters that can be tweaked grows quadratically with dimension, allowing more freedom for optimisation.


\inlineheading{Acknowledgements.}
This work was funded by Singapore National Research Foundation Fellowship NRF-NRFF2016-02, the Lee Kuan Yew Endowment Fund (Postdoctoral Fellowship), Singapore Ministry of Education Tier 1 grant RG190/17 and NRF-ANR grant NRF2017-NRF- ANR004 VanQuTe. Q.L., T.J.E.~and F.C.B.~thank the Centre for Quantum Technologies for their hospitality.

\bibliography{PhaseOptimization}

\begin{thebibliography}{33}%
\makeatletter
\providecommand \@ifxundefined [1]{%
 \@ifx{#1\undefined}
}%
\providecommand \@ifnum [1]{%
 \ifnum #1\expandafter \@firstoftwo
 \else \expandafter \@secondoftwo
 \fi
}%
\providecommand \@ifx [1]{%
 \ifx #1\expandafter \@firstoftwo
 \else \expandafter \@secondoftwo
 \fi
}%
\providecommand \natexlab [1]{#1}%
\providecommand \enquote  [1]{``#1''}%
\providecommand \bibnamefont  [1]{#1}%
\providecommand \bibfnamefont [1]{#1}%
\providecommand \citenamefont [1]{#1}%
\providecommand \href@noop [0]{\@secondoftwo}%
\providecommand \href [0]{\begingroup \@sanitize@url \@href}%
\providecommand \@href[1]{\@@startlink{#1}\@@href}%
\providecommand \@@href[1]{\endgroup#1\@@endlink}%
\providecommand \@sanitize@url [0]{\catcode `\\12\catcode `\$12\catcode
  `\&12\catcode `\#12\catcode `\^12\catcode `\_12\catcode `\%12\relax}%
\providecommand \@@startlink[1]{}%
\providecommand \@@endlink[0]{}%
\providecommand \url  [0]{\begingroup\@sanitize@url \@url }%
\providecommand \@url [1]{\endgroup\@href {#1}{\urlprefix }}%
\providecommand \urlprefix  [0]{URL }%
\providecommand \Eprint [0]{\href }%
\providecommand \doibase [0]{http://dx.doi.org/}%
\providecommand \selectlanguage [0]{\@gobble}%
\providecommand \bibinfo  [0]{\@secondoftwo}%
\providecommand \bibfield  [0]{\@secondoftwo}%
\providecommand \translation [1]{[#1]}%
\providecommand \BibitemOpen [0]{}%
\providecommand \bibitemStop [0]{}%
\providecommand \bibitemNoStop [0]{.\EOS\space}%
\providecommand \EOS [0]{\spacefactor3000\relax}%
\providecommand \BibitemShut  [1]{\csname bibitem#1\endcsname}%
\let\auto@bib@innerbib\@empty
\bibitem [{\citenamefont {Crutchfield}\ and\ \citenamefont
  {Young}(1989)}]{crutchfield1989inferring}%
  \BibitemOpen
  \bibfield  {author} {\bibinfo {author} {\bibfnamefont {J.~P.}\ \bibnamefont
  {Crutchfield}}\ and\ \bibinfo {author} {\bibfnamefont {K.}~\bibnamefont
  {Young}},\ }\href@noop {} {\bibfield  {journal} {\bibinfo  {journal}
  {Physical Review Letters}\ }\textbf {\bibinfo {volume} {63}},\ \bibinfo
  {pages} {105} (\bibinfo {year} {1989})}\BibitemShut {NoStop}%
\bibitem [{\citenamefont {Shalizi}\ and\ \citenamefont
  {Crutchfield}(2001)}]{shalizi2001computational}%
  \BibitemOpen
  \bibfield  {author} {\bibinfo {author} {\bibfnamefont {C.~R.}\ \bibnamefont
  {Shalizi}}\ and\ \bibinfo {author} {\bibfnamefont {J.~P.}\ \bibnamefont
  {Crutchfield}},\ }\href@noop {} {\bibfield  {journal} {\bibinfo  {journal}
  {Journal of Statistical Physics}\ }\textbf {\bibinfo {volume} {104}},\
  \bibinfo {pages} {817} (\bibinfo {year} {2001})}\BibitemShut {NoStop}%
\bibitem [{\citenamefont {Crutchfield}(2012)}]{crutchfield2012between}%
  \BibitemOpen
  \bibfield  {author} {\bibinfo {author} {\bibfnamefont {J.~P.}\ \bibnamefont
  {Crutchfield}},\ }\href@noop {} {\bibfield  {journal} {\bibinfo  {journal}
  {Nature Physics}\ }\textbf {\bibinfo {volume} {8}},\ \bibinfo {pages} {17}
  (\bibinfo {year} {2012})}\BibitemShut {NoStop}%
\bibitem [{\citenamefont {Crutchfield}\ and\ \citenamefont
  {Feldman}(1997)}]{crutchfield1997statistical}%
  \BibitemOpen
  \bibfield  {author} {\bibinfo {author} {\bibfnamefont {J.~P.}\ \bibnamefont
  {Crutchfield}}\ and\ \bibinfo {author} {\bibfnamefont {D.~P.}\ \bibnamefont
  {Feldman}},\ }\href@noop {} {\bibfield  {journal} {\bibinfo  {journal}
  {Physical Review E}\ }\textbf {\bibinfo {volume} {55}},\ \bibinfo {pages}
  {R1239} (\bibinfo {year} {1997})}\BibitemShut {NoStop}%
\bibitem [{\citenamefont {Palmer}\ \emph {et~al.}(2000)\citenamefont {Palmer},
  \citenamefont {Fairall},\ and\ \citenamefont
  {Brewer}}]{palmer2000complexity}%
  \BibitemOpen
  \bibfield  {author} {\bibinfo {author} {\bibfnamefont {A.~J.}\ \bibnamefont
  {Palmer}}, \bibinfo {author} {\bibfnamefont {C.~W.}\ \bibnamefont {Fairall}},
  \ and\ \bibinfo {author} {\bibfnamefont {W.~A.}\ \bibnamefont {Brewer}},\
  }\href@noop {} {\bibfield  {journal} {\bibinfo  {journal} {IEEE Transactions
  on Geoscience and Remote Sensing}\ }\textbf {\bibinfo {volume} {38}},\
  \bibinfo {pages} {2056} (\bibinfo {year} {2000})}\BibitemShut {NoStop}%
\bibitem [{\citenamefont {Varn}\ \emph {et~al.}(2002)\citenamefont {Varn},
  \citenamefont {Canright},\ and\ \citenamefont
  {Crutchfield}}]{varn2002discovering}%
  \BibitemOpen
  \bibfield  {author} {\bibinfo {author} {\bibfnamefont {D.~P.}\ \bibnamefont
  {Varn}}, \bibinfo {author} {\bibfnamefont {G.~S.}\ \bibnamefont {Canright}},
  \ and\ \bibinfo {author} {\bibfnamefont {J.~P.}\ \bibnamefont
  {Crutchfield}},\ }\href@noop {} {\bibfield  {journal} {\bibinfo  {journal}
  {Physical Review B}\ }\textbf {\bibinfo {volume} {66}},\ \bibinfo {pages}
  {174110} (\bibinfo {year} {2002})}\BibitemShut {NoStop}%
\bibitem [{\citenamefont {Clarke}\ \emph {et~al.}(2003)\citenamefont {Clarke},
  \citenamefont {Freeman},\ and\ \citenamefont
  {Watkins}}]{clarke2003application}%
  \BibitemOpen
  \bibfield  {author} {\bibinfo {author} {\bibfnamefont {R.~W.}\ \bibnamefont
  {Clarke}}, \bibinfo {author} {\bibfnamefont {M.~P.}\ \bibnamefont {Freeman}},
  \ and\ \bibinfo {author} {\bibfnamefont {N.~W.}\ \bibnamefont {Watkins}},\
  }\href@noop {} {\bibfield  {journal} {\bibinfo  {journal} {Physical Review
  E}\ }\textbf {\bibinfo {volume} {67}},\ \bibinfo {pages} {016203} (\bibinfo
  {year} {2003})}\BibitemShut {NoStop}%
\bibitem [{\citenamefont {Park}\ \emph {et~al.}(2007)\citenamefont {Park},
  \citenamefont {Lee}, \citenamefont {Yang}, \citenamefont {Jo},\ and\
  \citenamefont {Moon}}]{park2007complexity}%
  \BibitemOpen
  \bibfield  {author} {\bibinfo {author} {\bibfnamefont {J.~B.}\ \bibnamefont
  {Park}}, \bibinfo {author} {\bibfnamefont {J.~W.}\ \bibnamefont {Lee}},
  \bibinfo {author} {\bibfnamefont {J.-S.}\ \bibnamefont {Yang}}, \bibinfo
  {author} {\bibfnamefont {H.-H.}\ \bibnamefont {Jo}}, \ and\ \bibinfo {author}
  {\bibfnamefont {H.-T.}\ \bibnamefont {Moon}},\ }\href@noop {} {\bibfield
  {journal} {\bibinfo  {journal} {Physica A: Statistical Mechanics and its
  Applications}\ }\textbf {\bibinfo {volume} {379}},\ \bibinfo {pages} {179}
  (\bibinfo {year} {2007})}\BibitemShut {NoStop}%
\bibitem [{\citenamefont {Li}\ \emph {et~al.}(2008)\citenamefont {Li},
  \citenamefont {Yang},\ and\ \citenamefont {Komatsuzaki}}]{li2008multiscale}%
  \BibitemOpen
  \bibfield  {author} {\bibinfo {author} {\bibfnamefont {C.-B.}\ \bibnamefont
  {Li}}, \bibinfo {author} {\bibfnamefont {H.}~\bibnamefont {Yang}}, \ and\
  \bibinfo {author} {\bibfnamefont {T.}~\bibnamefont {Komatsuzaki}},\
  }\href@noop {} {\bibfield  {journal} {\bibinfo  {journal} {Proceedings of the
  National Academy of Sciences}\ }\textbf {\bibinfo {volume} {105}},\ \bibinfo
  {pages} {536} (\bibinfo {year} {2008})}\BibitemShut {NoStop}%
\bibitem [{\citenamefont {Haslinger}\ \emph {et~al.}(2010)\citenamefont
  {Haslinger}, \citenamefont {Klinkner},\ and\ \citenamefont
  {Shalizi}}]{haslinger2010computational}%
  \BibitemOpen
  \bibfield  {author} {\bibinfo {author} {\bibfnamefont {R.}~\bibnamefont
  {Haslinger}}, \bibinfo {author} {\bibfnamefont {K.~L.}\ \bibnamefont
  {Klinkner}}, \ and\ \bibinfo {author} {\bibfnamefont {C.~R.}\ \bibnamefont
  {Shalizi}},\ }\href@noop {} {\bibfield  {journal} {\bibinfo  {journal}
  {Neural Computation}\ }\textbf {\bibinfo {volume} {22}},\ \bibinfo {pages}
  {121} (\bibinfo {year} {2010})}\BibitemShut {NoStop}%
\bibitem [{\citenamefont {Kelly}\ \emph {et~al.}(2012)\citenamefont {Kelly},
  \citenamefont {Dillingham}, \citenamefont {Hudson},\ and\ \citenamefont
  {Wiesner}}]{kelly2012new}%
  \BibitemOpen
  \bibfield  {author} {\bibinfo {author} {\bibfnamefont {D.}~\bibnamefont
  {Kelly}}, \bibinfo {author} {\bibfnamefont {M.}~\bibnamefont {Dillingham}},
  \bibinfo {author} {\bibfnamefont {A.}~\bibnamefont {Hudson}}, \ and\ \bibinfo
  {author} {\bibfnamefont {K.}~\bibnamefont {Wiesner}},\ }\href@noop {}
  {\bibfield  {journal} {\bibinfo  {journal} {PloS one}\ }\textbf {\bibinfo
  {volume} {7}},\ \bibinfo {pages} {e29703} (\bibinfo {year}
  {2012})}\BibitemShut {NoStop}%
\bibitem [{\citenamefont {Gu}\ \emph {et~al.}(2012)\citenamefont {Gu},
  \citenamefont {Wiesner}, \citenamefont {Rieper},\ and\ \citenamefont
  {Vedral}}]{gu2012quantum}%
  \BibitemOpen
  \bibfield  {author} {\bibinfo {author} {\bibfnamefont {M.}~\bibnamefont
  {Gu}}, \bibinfo {author} {\bibfnamefont {K.}~\bibnamefont {Wiesner}},
  \bibinfo {author} {\bibfnamefont {E.}~\bibnamefont {Rieper}}, \ and\ \bibinfo
  {author} {\bibfnamefont {V.}~\bibnamefont {Vedral}},\ }\href@noop {}
  {\bibfield  {journal} {\bibinfo  {journal} {Nature Communications}\ }\textbf
  {\bibinfo {volume} {3}},\ \bibinfo {pages} {762} (\bibinfo {year}
  {2012})}\BibitemShut {NoStop}%
\bibitem [{\citenamefont {Mahoney}\ \emph {et~al.}(2016)\citenamefont
  {Mahoney}, \citenamefont {Aghamohammadi},\ and\ \citenamefont
  {Crutchfield}}]{mahoney2016occam}%
  \BibitemOpen
  \bibfield  {author} {\bibinfo {author} {\bibfnamefont {J.~R.}\ \bibnamefont
  {Mahoney}}, \bibinfo {author} {\bibfnamefont {C.}~\bibnamefont
  {Aghamohammadi}}, \ and\ \bibinfo {author} {\bibfnamefont {J.~P.}\
  \bibnamefont {Crutchfield}},\ }\href@noop {} {\bibfield  {journal} {\bibinfo
  {journal} {Scientific Reports}\ }\textbf {\bibinfo {volume} {6}},\ \bibinfo
  {pages} {20495} (\bibinfo {year} {2016})}\BibitemShut {NoStop}%
\bibitem [{\citenamefont {Thompson}\ \emph {et~al.}(2017)\citenamefont
  {Thompson}, \citenamefont {Garner}, \citenamefont {Vedral},\ and\
  \citenamefont {Gu}}]{thompson2017using}%
  \BibitemOpen
  \bibfield  {author} {\bibinfo {author} {\bibfnamefont {J.}~\bibnamefont
  {Thompson}}, \bibinfo {author} {\bibfnamefont {A.~J.~P.}\ \bibnamefont
  {Garner}}, \bibinfo {author} {\bibfnamefont {V.}~\bibnamefont {Vedral}}, \
  and\ \bibinfo {author} {\bibfnamefont {M.}~\bibnamefont {Gu}},\ }\href@noop
  {} {\bibfield  {journal} {\bibinfo  {journal} {npj Quantum Information}\
  }\textbf {\bibinfo {volume} {3}},\ \bibinfo {pages} {6} (\bibinfo {year}
  {2017})}\BibitemShut {NoStop}%
\bibitem [{\citenamefont {Aghamohammadi}\ \emph {et~al.}(2018)\citenamefont
  {Aghamohammadi}, \citenamefont {Loomis}, \citenamefont {Mahoney},\ and\
  \citenamefont {Crutchfield}}]{aghamohammadi2018extreme}%
  \BibitemOpen
  \bibfield  {author} {\bibinfo {author} {\bibfnamefont {C.}~\bibnamefont
  {Aghamohammadi}}, \bibinfo {author} {\bibfnamefont {S.~P.}\ \bibnamefont
  {Loomis}}, \bibinfo {author} {\bibfnamefont {J.~R.}\ \bibnamefont {Mahoney}},
  \ and\ \bibinfo {author} {\bibfnamefont {J.~P.}\ \bibnamefont
  {Crutchfield}},\ }\href@noop {} {\bibfield  {journal} {\bibinfo  {journal}
  {Physical Review X}\ }\textbf {\bibinfo {volume} {8}},\ \bibinfo {pages}
  {011025} (\bibinfo {year} {2018})}\BibitemShut {NoStop}%
\bibitem [{\citenamefont {Binder}\ \emph {et~al.}(2018)\citenamefont {Binder},
  \citenamefont {Thompson},\ and\ \citenamefont {Gu}}]{binder2018practical}%
  \BibitemOpen
  \bibfield  {author} {\bibinfo {author} {\bibfnamefont {F.~C.}\ \bibnamefont
  {Binder}}, \bibinfo {author} {\bibfnamefont {J.}~\bibnamefont {Thompson}}, \
  and\ \bibinfo {author} {\bibfnamefont {M.}~\bibnamefont {Gu}},\ }\href@noop
  {} {\bibfield  {journal} {\bibinfo  {journal} {Physical Review Letters}\
  }\textbf {\bibinfo {volume} {120}},\ \bibinfo {pages} {240502} (\bibinfo
  {year} {2018})}\BibitemShut {NoStop}%
\bibitem [{\citenamefont {Elliott}\ and\ \citenamefont
  {Gu}(2018)}]{elliott2018superior}%
  \BibitemOpen
  \bibfield  {author} {\bibinfo {author} {\bibfnamefont {T.~J.}\ \bibnamefont
  {Elliott}}\ and\ \bibinfo {author} {\bibfnamefont {M.}~\bibnamefont {Gu}},\
  }\href@noop {} {\bibfield  {journal} {\bibinfo  {journal} {npj Quantum
  Information}\ }\textbf {\bibinfo {volume} {4}},\ \bibinfo {pages} {18}
  (\bibinfo {year} {2018})}\BibitemShut {NoStop}%
\bibitem [{\citenamefont {Elliott}\ \emph {et~al.}(2018)\citenamefont
  {Elliott}, \citenamefont {Garner},\ and\ \citenamefont
  {Gu}}]{elliott2018quantum}%
  \BibitemOpen
  \bibfield  {author} {\bibinfo {author} {\bibfnamefont {T.~J.}\ \bibnamefont
  {Elliott}}, \bibinfo {author} {\bibfnamefont {A.~J.~P.}\ \bibnamefont
  {Garner}}, \ and\ \bibinfo {author} {\bibfnamefont {M.}~\bibnamefont {Gu}},\
  }\href@noop {} {\bibfield  {journal} {\bibinfo  {journal} {arXiv:1803.05426}\
  } (\bibinfo {year} {2018})}\BibitemShut {NoStop}%
\bibitem [{\citenamefont {Riechers}\ \emph {et~al.}(2016)\citenamefont
  {Riechers}, \citenamefont {Mahoney}, \citenamefont {Aghamohammadi},\ and\
  \citenamefont {Crutchfield}}]{riechers2016minimized}%
  \BibitemOpen
  \bibfield  {author} {\bibinfo {author} {\bibfnamefont {P.~M.}\ \bibnamefont
  {Riechers}}, \bibinfo {author} {\bibfnamefont {J.~R.}\ \bibnamefont
  {Mahoney}}, \bibinfo {author} {\bibfnamefont {C.}~\bibnamefont
  {Aghamohammadi}}, \ and\ \bibinfo {author} {\bibfnamefont {J.~P.}\
  \bibnamefont {Crutchfield}},\ }\href@noop {} {\bibfield  {journal} {\bibinfo
  {journal} {Physical Review A}\ }\textbf {\bibinfo {volume} {93}},\ \bibinfo
  {pages} {052317} (\bibinfo {year} {2016})}\BibitemShut {NoStop}%
\bibitem [{\citenamefont {Yang}\ \emph {et~al.}(2018)\citenamefont {Yang},
  \citenamefont {Binder}, \citenamefont {Narasimhachar},\ and\ \citenamefont
  {Gu}}]{yang2018matrix}%
  \BibitemOpen
  \bibfield  {author} {\bibinfo {author} {\bibfnamefont {C.}~\bibnamefont
  {Yang}}, \bibinfo {author} {\bibfnamefont {F.~C.}\ \bibnamefont {Binder}},
  \bibinfo {author} {\bibfnamefont {V.}~\bibnamefont {Narasimhachar}}, \ and\
  \bibinfo {author} {\bibfnamefont {M.}~\bibnamefont {Gu}},\ }\href@noop {}
  {\bibfield  {journal} {\bibinfo  {journal} {arXiv:1803.08220}\ } (\bibinfo
  {year} {2018})}\BibitemShut {NoStop}%
\bibitem [{\citenamefont {Palsson}\ \emph {et~al.}(2017)\citenamefont
  {Palsson}, \citenamefont {Gu}, \citenamefont {Ho}, \citenamefont {Wiseman},\
  and\ \citenamefont {Pryde}}]{palsson2017experimentally}%
  \BibitemOpen
  \bibfield  {author} {\bibinfo {author} {\bibfnamefont {M.~S.}\ \bibnamefont
  {Palsson}}, \bibinfo {author} {\bibfnamefont {M.}~\bibnamefont {Gu}},
  \bibinfo {author} {\bibfnamefont {J.}~\bibnamefont {Ho}}, \bibinfo {author}
  {\bibfnamefont {H.~M.}\ \bibnamefont {Wiseman}}, \ and\ \bibinfo {author}
  {\bibfnamefont {G.~J.}\ \bibnamefont {Pryde}},\ }\href@noop {} {\bibfield
  {journal} {\bibinfo  {journal} {Science Advances}\ }\textbf {\bibinfo
  {volume} {3}},\ \bibinfo {pages} {e1601302} (\bibinfo {year}
  {2017})}\BibitemShut {NoStop}%
\bibitem [{\citenamefont {Ghafari~Jouneghani}\ \emph
  {et~al.}(2017)\citenamefont {Ghafari~Jouneghani}, \citenamefont {Gu},
  \citenamefont {Ho}, \citenamefont {Thompson}, \citenamefont {Suen},
  \citenamefont {Wiseman},\ and\ \citenamefont {Pryde}}]{ghafari2017observing}%
  \BibitemOpen
  \bibfield  {author} {\bibinfo {author} {\bibfnamefont {F.}~\bibnamefont
  {Ghafari~Jouneghani}}, \bibinfo {author} {\bibfnamefont {M.}~\bibnamefont
  {Gu}}, \bibinfo {author} {\bibfnamefont {J.}~\bibnamefont {Ho}}, \bibinfo
  {author} {\bibfnamefont {J.}~\bibnamefont {Thompson}}, \bibinfo {author}
  {\bibfnamefont {W.~Y.}\ \bibnamefont {Suen}}, \bibinfo {author}
  {\bibfnamefont {H.~M.}\ \bibnamefont {Wiseman}}, \ and\ \bibinfo {author}
  {\bibfnamefont {G.~J.}\ \bibnamefont {Pryde}},\ }\href@noop {} {\bibfield
  {journal} {\bibinfo  {journal} {arXiv:1711.03661}\ } (\bibinfo {year}
  {2017})}\BibitemShut {NoStop}%
\bibitem [{\citenamefont {Garner}\ \emph {et~al.}(2017)\citenamefont {Garner},
  \citenamefont {Liu}, \citenamefont {Thompson}, \citenamefont {Vedral} \emph
  {et~al.}}]{garner2017provably}%
  \BibitemOpen
  \bibfield  {author} {\bibinfo {author} {\bibfnamefont {A.~J.~P.}\
  \bibnamefont {Garner}}, \bibinfo {author} {\bibfnamefont {Q.}~\bibnamefont
  {Liu}}, \bibinfo {author} {\bibfnamefont {J.}~\bibnamefont {Thompson}},
  \bibinfo {author} {\bibfnamefont {V.}~\bibnamefont {Vedral}},  \emph
  {et~al.},\ }\href@noop {} {\bibfield  {journal} {\bibinfo  {journal} {New
  Journal of Physics}\ }\textbf {\bibinfo {volume} {19}},\ \bibinfo {pages}
  {103009} (\bibinfo {year} {2017})}\BibitemShut {NoStop}%
\bibitem [{\citenamefont {Aghamohammadi}\ \emph {et~al.}(2017)\citenamefont
  {Aghamohammadi}, \citenamefont {Mahoney},\ and\ \citenamefont
  {Crutchfield}}]{aghamohammadi2017extreme}%
  \BibitemOpen
  \bibfield  {author} {\bibinfo {author} {\bibfnamefont {C.}~\bibnamefont
  {Aghamohammadi}}, \bibinfo {author} {\bibfnamefont {J.~R.}\ \bibnamefont
  {Mahoney}}, \ and\ \bibinfo {author} {\bibfnamefont {J.~P.}\ \bibnamefont
  {Crutchfield}},\ }\href@noop {} {\bibfield  {journal} {\bibinfo  {journal}
  {Scientific Reports}\ }\textbf {\bibinfo {volume} {7}} (\bibinfo {year}
  {2017})}\BibitemShut {NoStop}%
\bibitem [{\citenamefont {Thompson}\ \emph {et~al.}(2018)\citenamefont
  {Thompson}, \citenamefont {Garner}, \citenamefont {Mahoney}, \citenamefont
  {Crutchfield}, \citenamefont {Vedral},\ and\ \citenamefont
  {Gu}}]{thompson2018causal}%
  \BibitemOpen
  \bibfield  {author} {\bibinfo {author} {\bibfnamefont {J.}~\bibnamefont
  {Thompson}}, \bibinfo {author} {\bibfnamefont {A.~J.~P.}\ \bibnamefont
  {Garner}}, \bibinfo {author} {\bibfnamefont {J.~R.}\ \bibnamefont {Mahoney}},
  \bibinfo {author} {\bibfnamefont {J.~P.}\ \bibnamefont {Crutchfield}},
  \bibinfo {author} {\bibfnamefont {V.}~\bibnamefont {Vedral}}, \ and\ \bibinfo
  {author} {\bibfnamefont {M.}~\bibnamefont {Gu}},\ }\href@noop {} {\bibfield
  {journal} {\bibinfo  {journal} {Physical Review X}\ }\textbf {\bibinfo
  {volume} {8}},\ \bibinfo {pages} {031013} (\bibinfo {year}
  {2018})}\BibitemShut {NoStop}%
\bibitem [{\citenamefont {Suen}\ \emph {et~al.}(2017)\citenamefont {Suen},
  \citenamefont {Thompson}, \citenamefont {Garner}, \citenamefont {Vedral},\
  and\ \citenamefont {Gu}}]{suen2017classical}%
  \BibitemOpen
  \bibfield  {author} {\bibinfo {author} {\bibfnamefont {W.~Y.}\ \bibnamefont
  {Suen}}, \bibinfo {author} {\bibfnamefont {J.}~\bibnamefont {Thompson}},
  \bibinfo {author} {\bibfnamefont {A.~J.~P.}\ \bibnamefont {Garner}}, \bibinfo
  {author} {\bibfnamefont {V.}~\bibnamefont {Vedral}}, \ and\ \bibinfo {author}
  {\bibfnamefont {M.}~\bibnamefont {Gu}},\ }\href@noop {} {\bibfield  {journal}
  {\bibinfo  {journal} {Quantum}\ }\textbf {\bibinfo {volume} {1}},\ \bibinfo
  {pages} {25} (\bibinfo {year} {2017})}\BibitemShut {NoStop}%
\bibitem [{\citenamefont {Aghamohammadi}\ \emph {et~al.}(2016)\citenamefont
  {Aghamohammadi}, \citenamefont {Mahoney},\ and\ \citenamefont
  {Crutchfield}}]{aghamohammadi2016ambiguity}%
  \BibitemOpen
  \bibfield  {author} {\bibinfo {author} {\bibfnamefont {C.}~\bibnamefont
  {Aghamohammadi}}, \bibinfo {author} {\bibfnamefont {J.~R.}\ \bibnamefont
  {Mahoney}}, \ and\ \bibinfo {author} {\bibfnamefont {J.~P.}\ \bibnamefont
  {Crutchfield}},\ }\href@noop {} {\bibfield  {journal} {\bibinfo  {journal}
  {Physics Letters A}\ }\textbf {\bibinfo {volume} {381}},\ \bibinfo {pages}
  {1223} (\bibinfo {year} {2016})}\BibitemShut {NoStop}%
\bibitem [{\citenamefont {Loomis}\ and\ \citenamefont
  {Crutchfield}(2018)}]{loomis2018strong}%
  \BibitemOpen
  \bibfield  {author} {\bibinfo {author} {\bibfnamefont {S.}~\bibnamefont
  {Loomis}}\ and\ \bibinfo {author} {\bibfnamefont {J.~P.}\ \bibnamefont
  {Crutchfield}},\ }\href@noop {} {\bibfield  {journal} {\bibinfo  {journal}
  {arXiv:1808.08639}\ } (\bibinfo {year} {2018})}\BibitemShut {NoStop}%
\bibitem [{\citenamefont
  {Khintchine}(1934)}]{khintchine1934korrelationstheorie}%
  \BibitemOpen
  \bibfield  {author} {\bibinfo {author} {\bibfnamefont {A.}~\bibnamefont
  {Khintchine}},\ }\href@noop {} {\bibfield  {journal} {\bibinfo  {journal}
  {Mathematische Annalen}\ }\textbf {\bibinfo {volume} {109}},\ \bibinfo
  {pages} {604} (\bibinfo {year} {1934})}\BibitemShut {NoStop}%
\bibitem [{Note1()}]{Note1}%
  \BibitemOpen
  \bibinfo {note} {Note that $C_q$ has sometimes been alternatively referred to
  as the quantum statistical complexity or quantum machine complexity; we avoid
  such nomenclature here as the former is incorrect if the model is not
  minimal, while the latter invites potential confusion with
  $D_q$.}\BibitemShut {Stop}%
\bibitem [{\citenamefont {Nielsen}\ and\ \citenamefont
  {Chuang}(2000)}]{nielsen2000quantum}%
  \BibitemOpen
  \bibfield  {author} {\bibinfo {author} {\bibfnamefont {M.~A.}\ \bibnamefont
  {Nielsen}}\ and\ \bibinfo {author} {\bibfnamefont {I.}~\bibnamefont
  {Chuang}},\ }\href@noop {} {\enquote {\bibinfo {title} {Quantum {C}omputation
  and {Q}uantum {I}nformation},}\ } (\bibinfo {year} {2000})\BibitemShut
  {NoStop}%
\bibitem [{\citenamefont {Jozsa}\ and\ \citenamefont
  {Schlienz}(2000)}]{jozsa2000distinguishability}%
  \BibitemOpen
  \bibfield  {author} {\bibinfo {author} {\bibfnamefont {R.}~\bibnamefont
  {Jozsa}}\ and\ \bibinfo {author} {\bibfnamefont {J.}~\bibnamefont
  {Schlienz}},\ }\href@noop {} {\bibfield  {journal} {\bibinfo  {journal}
  {Physical Review A}\ }\textbf {\bibinfo {volume} {62}},\ \bibinfo {pages}
  {012301} (\bibinfo {year} {2000})}\BibitemShut {NoStop}%
\bibitem [{\citenamefont {Horodecki}\ and\ \citenamefont
  {Oppenheim}(2013)}]{horodecki2013fundamental}%
  \BibitemOpen
  \bibfield  {author} {\bibinfo {author} {\bibfnamefont {M.}~\bibnamefont
  {Horodecki}}\ and\ \bibinfo {author} {\bibfnamefont {J.}~\bibnamefont
  {Oppenheim}},\ }\href@noop {} {\bibfield  {journal} {\bibinfo  {journal}
  {Nature Communications}\ }\textbf {\bibinfo {volume} {4}},\ \bibinfo {pages}
  {2059} (\bibinfo {year} {2013})}\BibitemShut {NoStop}%
\end{thebibliography}%

\clearpage

\widetext
\section*{Supplementary Material}
\setcounter{equation}{0}
\setcounter{figure}{0}
\setcounter{table}{0}
\setcounter{page}{1}
\renewcommand{\theequation}{S\arabic{equation}}
\renewcommand{\thefigure}{S\arabic{figure}}
\renewcommand{\thepage}{S\arabic{page}}

\begin{center}
\textbf{Supplementary A: Existence of $\uphase$ and overlap of quantum memory states}
\end{center}

Here we show that the unitary operator $\uphase$ for our phase-encoded quantum models exists for any choice of the phases $\varphi_{xj}$, provide an expression for the overlaps of pairs of quantum memory states, and show that the solution to this overlap converges.

\inlineheading{Existence of $\uphase$.}
We introduce the notation $\onephase{j}$ to indicate the combined system-ancilla state after applying the unitary circuit:
\begin{equation}
\onephase{j}:=\uphase\sigmaphase{j}\ket{0}=\sum_{x}\sqrt{P(x|j)}e^{i\varphi_{xj}}\sigmaphase{\lambda(j,x)}\ket{x}.
\end{equation}
Previous work~\cite{binder2018practical} established the existence of a unitary operation $U$ in the non-phase-encoded case if and only if
\begin{equation}
\braket{\sigma_j}{\sigma_k}=\braket{1_j}{1_k} \;\forall\ j,k.
\end{equation}
Similarly, for the existence of $\uphase$ in our phase-encoded models we require:
\begin{align}
\label{eq:1i1j}
\braketsigmaphase{j}{k}=\braketonephase{j}{k}&=\sum_{x}\sqrt{P(x|j)}e^{-i\varphi_{xj}}\brasigmaphase{\lambda(j,x)}\bra{x} \sum_{x'} \sqrt{P(x'|k)}e^{i\varphi_{x'k}}\sigmaphase{\lambda(k,x')}\ket{x'}\nonumber\\
&=\sum_{x}\sqrt{P(x|j)P(x|k)}e^{i(\varphi_{xk}-\varphi_{xj})}\braketsigmaphase{\lambda(j,x)}{\lambda(k,x)}.
\end{align}
A solution for the inner product of the quantum memory states is as follows:
\begin{equation}\label{eq:cij}
\cijphase{j}{k}:=\braketsigmaphase{j}{k}=\sum_{\future{x}} \sqrt{P(\future{x}|j)P(\future{x}|k)}e^{i(\varphi_{\future{x}k}-\varphi_{\future{x}j})},
\end{equation}
which can be verified by insertion into Eq.~\eqref{eq:1i1j}, thus proving the existence of $\uphase$.

\inlineheading{Convergence of $\cijphase{j}{k}$.}
We must now verify that our solution to the memory state overlaps is convergent; that is, $\lim_{L\to\infty} \cijphase{j}{k}^{[L]}=\cijphase{j}{k}$, where
\begin{equation}
\label{eq:conv}
\cijphase{j}{k}^{[L]}:=\sum_{x_{0:L}} \sqrt{P(X_{0:L}=x_{0:L}|S_0=j)P(X_{0:L}=x_{0:L}|S_0=k)}e^{i(\varphi_{x_{0:L}k}-\varphi_{x_{0:L}j})}\cijphase{\lambda(j,x_{0:L})}{\lambda(k,x_{0:L})}.
\end{equation}
Note that to avoid confusion between variables at different timesteps, in this section we do not employ the shorthand $P(x|j)$ introduced in the main text.

We assume that we are dealing with synchronisable processes, such that the memory of the model can be initialised properly given the entire past. Recalling that $H(A|B):=\sum_bP(B=b)H(A|B=b)$, this condition can be expressed
\begin{equation}\label{eq:h}
\lim_{L\to\infty}H(S_0|X_{-L:0})=0,
\end{equation}
and thus for large $L$ we can express
\begin{equation}\label{eq:pxl}
\sum_{x_{-L:0}}P(X_{-L:0}=x_{-L:0})H\left(S_0|X_{-L:0}=x_{-L:0}\right)<\varepsilon(L)
\end{equation}
for some small $\varepsilon(L)$ that vanishes as $L\to\infty$. This allows us to divide the possible trajectories $x_{-L:0}$ into two classes: those where the memory state is (asymptotically) synchronised ($H(S_0|\past{x})=0$); and those where it is not. However, since this uncertainty is finite, the probability of such non-synchronising trajectories occuring must be vanishingly small for consistency with Eq.~\eqref{eq:pxl}, and moreover, the total probability of such trajectories must also be vanishingly small. We can therefore devote our attention only to the former class.

For this former class, we can express
\begin{equation}\label{eq:pkxl}
H\left(S_0|X_{-L:0}=x_{-L:0}\right)=-\sum_{j}P(S_0=j|X_{-L:0}=x_{-L:0})\log_2 (P(S_0=j|X_{-L:0}=x_{-L:0}))< \varepsilon'(L)
\end{equation}
for some $\varepsilon'(L)$ that again vanishes as $L\to\infty$. Since each term in the summation is non-negative, we can also constrain each term to satisfy the inequality individually. To satisfy this, we must have that each $P(S_0=j|X_{-L:0}=x_{0:L})$ is either close to 0 or 1. These probabilities must sum to 1, which ensures that for one value of $s_0$, which we shall label as $m$, the probability is $1-\epsilon(L)$ for some small $\epsilon(L)$, while the others occur with probability $\epsilon_j$ that are each also small, with $\sum_{j\neq m}\epsilon _j(L)=\epsilon(L)$. In other words, after having produced a sufficiently long sequence of outputs $x_{-L:0}$ the past of the process almost certainly belongs to causal state $m$, and
\begin{equation}
\lim_{L\to \infty}P(S_0=j|X_{-L:0}=x_{-L:0})=\delta_{jm}.
\end{equation}
Now consider the expansion
\begin{align}
P(S_0=k|X_{-L:0}=x_{-L:0})&=\sum_jP(S_0=k, S_{-L}=j|X_{-L:0}=x_{-L:0})\nonumber\\
&=\sum_{j}P(S_{-L}=j|X_{-L:0}=x_{-L:0})P(S_0=k|S_{-L}=j,X_{-L:0}=x_{-L:0})\nonumber\\
&=\sum_{j}P(S_{-L}=j|X_{-L:0}=x_{-L:0})\delta_{k\lambda(j,x_{-L:0})}.
\end{align}
For $L\to\infty$, the left-hand side becomes arbitrarily close to 1 when $k=m$, and 0 otherwise.

Examining the case $k=m$, since $\sum_{j}P(S_{-L}=j|X_{-L:0}=x_{-L:0})=1$, for any $j$ where $k\neq\lambda(j,x_{-L:0})$ we must have $P(S_{-L}=j|X_{-L:0}=x_{-L:0})\approx 0$. Using Bayes' rule, and assuming that $P(S=j)\not\approx 0$, we have
\begin{equation}
\frac{P(X_{-L:0}=x_{-L:0}|S_{-L}=j)}{P(X_{-L:0}=x_{-L:0})}\approx0
\end{equation}
implying that for any $j$ such that $k\neq\lambda(j,x_{-L:0})$, the probability of such an output trajectory occuring given we started in a past belonging to causal state $j$ must be vanishingly small, even relative to the probability of the trajectory occuring at all.

Taken together, these lead us to the conclusion that
\begin{equation}
\lim_{L\to\infty} \lambda(j,x_{-L:0})=\lim_{L\to\infty}\lambda(x_{-L:0})
\end{equation}
for all but a set of output trajectories of vanishingly small probability; that is, for sufficiently large $L$ the current causal state is almost certainly determined by the output sequence alone independent of the initial state prior to this sequence. Note that for processes with finite Markov order this statement is tautologically true for any trajectory once $L$ is at least as large as the Markov order.

Returning then to Eq.~\eqref{eq:conv}, we see that for sufficiently large $L$ that for all but a set of trajectories of vanishingly small probability we may replace $\cijphase{\lambda(j,x_{0:L})}{\lambda(k,x_{0:L})}\to\cijphase{\lambda(x_{0:L})}{\lambda(x_{0:L})}=1$. Thus, for sufficiently large $L$, the recursive factor in the expression tends towards unity, and as such $\lim_{L\to\infty} \cijphase{j}{k}^{[L]}=\cijphase{j}{k}$ as required.


\begin{center}
\textbf{Supplementary B. Numerical sweep search for phase-enhancements}
\end{center}

\begin{figure}
\begin{center}
\includegraphics[width=0.5\linewidth]{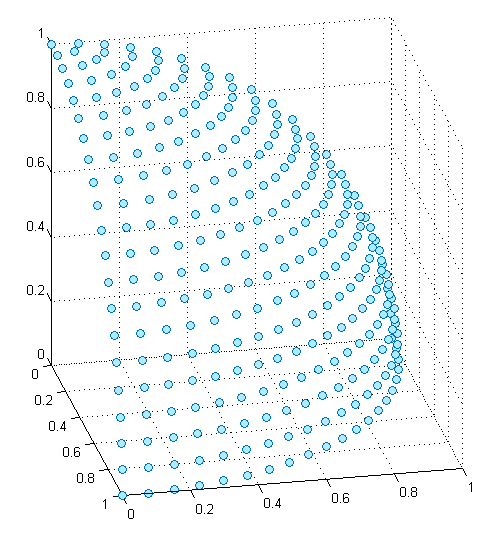}
\end{center}
\caption{The transition probabilities out of each state in a three-state Markov process can be represented as a point on the positive octant of a unit sphere. Three such points define the process. We sweep over a discretised set of such points in our numerical treatment to systematically study these processes.}
\label{fig:sweep}
\end{figure}

For a general three-state Markov process as depicted in \figref{fig:general_process}, each state is described by the three output probabilities to each state, defined by two free parameters due to normalisation of probability. These free parameters can be mapped to a point on the positive octant of a unit sphere [\figref{fig:sweep}], where the square of the distance along a given axis corresponds to the probability of transitioning into the corresponding state. Each process is defined by three such points, one for each state.

In the case of searching for dimensional advantages, we systematically sweep over these surfaces, coarse-grained into grids such that there are 20 evenly-spaced steps along each edge of the sweep areas. For each process we then check whether the inequalities Eqs.~\eqref{eq:inequality} are satisfied for any of the combinations of $\alpha$ and $\beta$ given in the main text.

When searching for entropic advantages, we instead sample by randomly picking a point on the three surfaces to determine a process, and then systematically sweep over all possible phase angles for the process to seek whether a $\cphase<C_q$ can be found.

Our findings are summarised in the table below:

\vspace{1em}
\begin{center}
\begin{tabular}{|l|c|}
\hline
\textbf{Advantage}                      & \textbf{$\%$ of three-state processes admitting advantage} \\ \hline
Entropic                                & $<0.5$                                                          \\
Dimensional ($\alpha=\beta=1$)          & $\sim9$                                                         \\
Dimensional (Multiple ($\alpha,\beta$)) & $>17$                                                           \\ \hline
\end{tabular}
\end{center}

\end{document}